\documentclass[runningheads]{llncs}
\usepackage{graphicx}
\begin{document}
\title{Audit and Assurance of AI Algorithms: A framework to ensure ethical algorithmic practices in Artificial Intelligence}
%
%
\author{Ramya Akula\inst{1}\orcidID{0000-0002-4008-075X} \and
Ivan Garibay\inst{1}\orcidID{0000-0002-3302-9382}}
\authorrunning{R. Akula et al.}
%
\institute{University of Central Florida, USA\\
\email{\{ramya.akula@knights.ucf.edu, igaribay@ucf.edu\}}\\
}
\maketitle              
\begin{abstract}
	Algorithms are becoming more widely used in business, and businesses are becoming increasingly concerned that their algorithms will cause significant reputational or financial damage. We should emphasize that any of these damages stem from situations in which the United States lacks strict legislative prohibitions or specified protocols for measuring damages. As a result, governments are enacting legislation and enforcing prohibitions, regulators are fining businesses, and the judiciary is debating whether or not to make artificially intelligent computer models as the decision-makers in the eyes of the law. From autonomous vehicles and banking to medical care, housing, and legal decisions, there will soon be enormous amounts of algorithms that make decisions with limited human interference. Governments, businesses, and society would have an algorithm audit, which would have systematic verification that algorithms are lawful, ethical, and secure, similar to financial audits. A modern market, auditing, and assurance of algorithms developed to professionalize and industrialize AI, machine learning, and related algorithms. Stakeholders of this emerging field include policymakers and regulators, along with industry experts and entrepreneurs. In addition, we foresee audit thresholds and frameworks providing valuable information to all who are concerned with governance and standardization. This paper aims to review the critical areas required for auditing and assurance and spark discussion in this novel field of study and practice.

\keywords{Audit  \and Assurance \and Artificial Intelligence.}
\end{abstract}

\section{Introduction}
Rise of Artificial Intelligence(AI) and Machine Learning(ML)in various sectors such as law enforcement, health care, and banking concerns the safety, legality, and ethical consequences of its usage. We are currently in an era of the AI revolution, where confidentiality, ownership, accountability, and safety of algorithms are increasingly becoming a top priority. As AI matures, there will soon be hundreds of millions of algorithms making crucial decisions with little human intervention. It increases the need for frameworks that help audit the integrity of such algorithms concerning reliability, legality, fairness, and regulatory compliance. The availability of a framework that can audit the integrity of AI systems will increase the adoption of AI to new sectors in the different industries and help reduce production costs and increased revenue streams \cite{bellotti2020predicting}. While the previous decade focus on information security, the current emphasis is on algorithm integrity. Building a framework for checking the integrity of AI algorithms requires the development of new technologies, processes, and standards with inputs from government, industry, and society. This framework development provides an opportunity to alleviate the current concerns with the application of AI. Algorithm Auditing is the science and practice of evaluating, mitigating, and ensuring algorithms' safety, legality, and ethicality. It requires cutting-edge research in AI towards fairness, explainability, reliability, privacy, and classical issues such as data ethics \cite{goldsteen2020data}, administration, and governance. Akin to financial audits, ultimately, governments, industry, and society will also need algorithm audits, which is the formal guarantee that algorithms are legal, ethical, and safe.

\begin{figure*}
    \centering
    \includegraphics[scale=0.32] {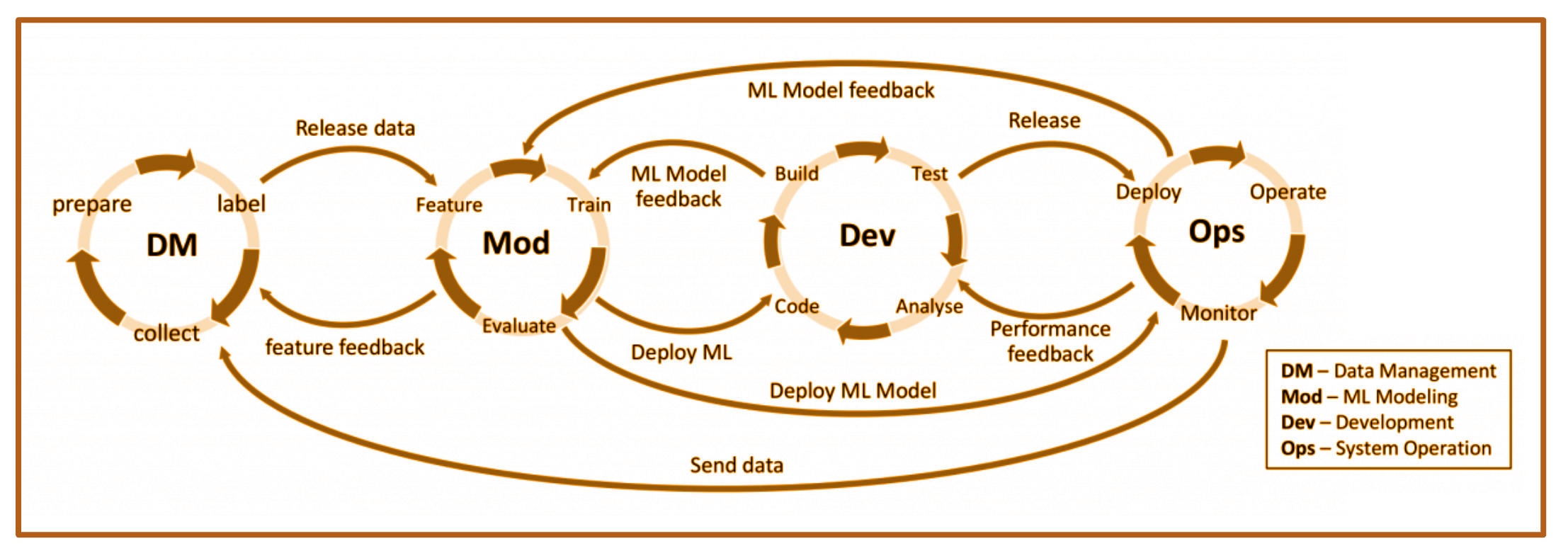}
    \caption{Four phases of AI application development: Data Management, Model Selection, Development, and Operation.}
    \label{fig:mlselifecycle} 
\end{figure*}

\section{Why Algorithm Audit}
In this section, we describe the components which constitute an algorithm auditing framework.
\subsection{Algorithms}
An algorithm is a set of well-defined sequences of instructions that are required to solve a problem. Expert systems are the first generation of AI algorithms, developed in the 1980s and 1990s, with many applications in the health care, finance, and manufacturing sectors \cite{smietanka2021algorithms}. Despite many efforts put into the research and prototyping of these systems, these were not very successful due to high operational costs. As shown in Figure \ref{fig:mlselifecycle}, the development of ML models regardless of the application includes four phases:
\begin{itemize}
    \item \textbf{Data Management}: This step includes collection, storage, extraction, normalization, transformation, and loading of data to assure well-structured data pipelines. This step assures that the ML task is well-defined and planned, along with documentation of data and software artifacts. It also includes selecting, refining, modifying, and structuring a feature space.
    \item \textbf{Model Selection}: Cross-validation, optimization, and comparison of models are all part of the model selection process.
    \item \textbf{development}: Enhances interpretability by adding thresholds, additional tools, and feedback mechanisms, presenting findings to key stakeholders, and assessing the algorithmic system's effect on the organization.
    \item \textbf{Operation}: Formulating and implementing supervision and supply interfaces after going through various review procedures, from IT to Business, keeping a proper record of outcomes and comments received in the field.
\end{itemize}
Even though these phases seem to be stable and self-contained, they interact compellingly, following a series of loops rather than a linear development. Though most of the research on each step happens in isolation, the scientific and technological communities are increasingly aware of the knowledge exchange and relationships. Each of these phases can be audited independently on the way to auditing the complete algorithm. Designers of the algorithm lay out a policy document ahead of time, stating what the algorithm intends to accomplish, making up the declaration of purpose while easing the audit.

\subsection{Accessibility}
The degree of access available to the auditing framework for a given algorithm may vary at different phases. In the typical research literature, the algorithms divide into two categories: 'White-box' and 'Black-box.' However, in practice, there are algorithms that are in between these categories along with multiple "shades of gray." As a result, there are potentially several degrees of system access for auditors. The highest degree is the white-box, where all the algorithm components, including input data, model architecture, training parameters, are available for audit.  The intermediate levels include access to the model or the portion of training data. The lowest degree is the black box, where there is no access to the model or the training data.   

\subsection{Audit's Outcome}
An auditing procedure's overall goal is to boost confidence or guarantee trust in the underlying system, which may subsequently document via a certification process. Based on the audit outcome, measures to reduce the risk interventions may enhance the result of the system across the different phases of the algorithm development. This mitigation approach will be increasingly focused, technological, and varied. The auditing process determines if the system complies with regulatory, governance, and ethical requirements once it assesses and implements mitigation measures \cite{larsson2020governance}. As a result, a worthwhile assurance process includes certification, governance, general and sector-specific assurance, insurance, monitoring interfaces, and predicting unknown risks.

\section{Algorithm Audit Phases}
\begin{figure*}[h!]
    \centering
    \includegraphics[scale=0.25] {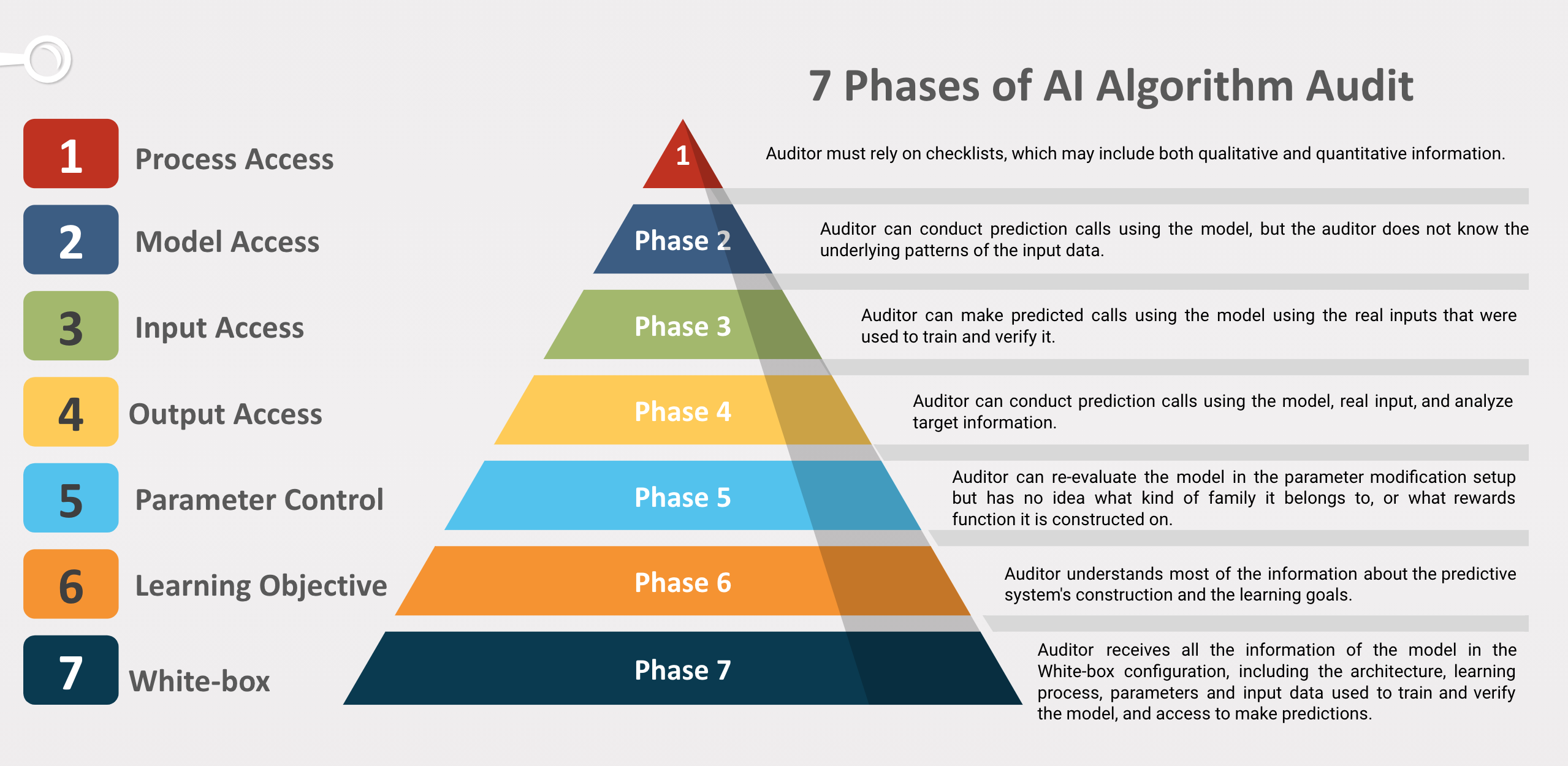}
    \caption{Seven potential phases for the AI Algorithm Audit. In each phase, an auditor has various degrees of access to conduct legitimate check.}
    \label{fig:phases} 
\end{figure*}

The need for AI Algorithm Audit to be repeatable and scalable is critical. During the AI Algorithm Audit, an auditor has various degrees of access. In reality, the knowledge spectrum of technology is more likely to be 'shades of grey,' i.e., a continuum, rather than an apparent dichotomy. This added complexity allows for a more in-depth examination of the technological innovations for vulnerability assessment and the appropriate degree of transparency. Audit varies from process-access to white-box, as shown in Figure \ref{fig:phases}.

\subsection{Audit Phase 1: Process Access}
The auditor has no direct access to the algorithm in the Process access configuration. Therefore the inspections and modifications take place throughout the model building process. Due to the inability to debug the model, the auditor must rely on checklists, including qualitative and quantitative information. The body of the evaluation may consist of general, and sector-specific rules provided by regulators and other governmental organizations, augmented with a combination of corporation recommendations \cite{hagendorff2020ethics}. This degree of transparency and feedback depth may be the most suitable for low-stakes and low-risk applications. When risks are minimal, monitoring is required and uses a checklist-phase approach. When the hazards are minimal to medium, and no monitoring is required, it is a black-box phase. When the risks are moderate, and some monitoring is required, uses the grey-box phase. When the dangers are medium-high and complete supervision is required, uses the glass-box phase.

\subsection{Audit Phase 2: Model Access}
In this phase, the auditor can conduct prediction calls using the model, but the auditor does not know the underlying patterns of the input data. Some information, such as the names of the parameters, types, and ranges, may be shared. As a result, the auditor can only make calls using some fictitious input. Because no data sharing agreements are required, this phase of access reduces the amount of information given to the auditor. The only access to the application programming interface is required to conduct the analysis and accomplish a high degree of automation at this phase.

\subsection{Audit Phase 3: Input Access}
In this phase, the auditor can make predicted calls using the model using the actual inputs used to train and verify it, but they cannot compare the predictions to the actual result data. The auditor alone has access to the input data needed to train and verify the model and the ability to make predictions. The lack of result data makes it challenging to evaluate a model's generalization behavior, much alone its performance. Some analysis is needed because just the predictions are provided, such as calculating bias from the uniformity of result, property and participation inference, or surrogate explanation. Creating synthetic data that is close to the actual distribution of the input allows for an audit of the model's brittleness to incremental changes in the distribution.

\subsection{Audit Phase 4: Output Access}
In this phase, the auditor can conduct prediction calls using the model, real input, and analyze target information. As a result, the auditor gets access to the output and input data used to train and verify the model and the ability to make predictions. From a modeling standpoint, there are various methods for assessing and operating at this phase, the majority of which fall under the category of model-agnostic processes. The auditor may conduct concept drift analysis, examine the correctness of explanations, execute inversion attacks, and verify bias from an equality of opportunity standpoint using the available access and a few assumptions. In addition, the auditor may create a baseline or alternative models to the original.

\subsection{Audit Phase 5: Parameter Control}
The auditor can re-evaluate the model in the parameter modification setup but has no idea what kind of family it belongs to or what rewards function it constructs on. The auditor has admin rights to the model's parameters, output, input data, and the ability to make predictions. The auditor may conduct explicit consistency and perturbations testing on the model at this phase. Thus, it is possible to offer good feedback, especially regarding the system's stability, its judgments, and the explanations supplied. It would also enable the auditor to evaluate the risk of operational theft from a privacy standpoint. This phase of access is simple to set up using an API, and it automates for external audits. Due to the lack of enough information on the model nature, there is little risk of infringement of intellectual property or other types of disclosures. In addition, since the auditor may re-parametrize the model and retrain it based on specific hypotheses, the auditor can retrain the model in practice.

\subsection{Audit Phase 6: Learning objective}
The auditor understands most of the information about the predictive system's construction and the learning goals. Auditor has access to parameters, output, and input data needed to train and verify the model and make predictions. The auditor understands how to refit the model using the actual objective function of training. The feedback is very detailed, with information on network size, stress-testing, and trade-off analysis of bias, privacy, and loss, all possible without making any assumptions. Because the human participation after setting up the interfaces and environments is minimal, this phase of access is sufficient for automated internal and external audits.

\subsection{Phase 7: White-box}
The auditor receives all the information of the model in the White-box configuration, including the architecture or type, learning process, task goals, parameters, output and input data used to train and verify the model, and access to make predictions. This degree of access, which is very similar to what the development team and potential customer have, enables the auditor to give more accurate and detailed comments. It would be simpler to evaluate mitigation measures and give accurate data that developers could more readily record. This degree of access is better suited to internal auditors or in-house consultants since it requires greater transparency, including information sharing and other agreements.

\section{Audit Assurance}
An auditing process's overall goal is to boost confidence in or guarantee trust in the underlying system. The auditing process determines if the system meets regulatory, governance, and ethical requirements after evaluating it and adopting mitigating measures. As a result, providing assurance must be understood in many dimensions and measures to demonstrate the algorithm's trustworthiness. With an increased application of AI in different sectors, compliance with a certain standard such as certification and continuous audit becomes mandatory. These mandatory standards can be either general or industry-specific. General standards are the guidelines that cover essential categories such as privacy, explainability, safety, and fairness by bodies and agencies with non-sector-specific remits. The progress in this area is becoming more apparent. Sector-specific standards are already available as many sectors are establishing their respective standards and methods to best practice, in addition to those provided by sector-specific authorities. There are also some application-specific standards, such as Facial Recognition used by law enforcement.

After setting up standards, the next important step is to make sure that they are followed by setting up an administrative body. The governing body would deal with both non-technical and technical aspects. Non-technical governance refers to the structures and procedures responsible for assigning decision-makers, providing proper training and skills, keeping the human in the loop, and performing environmental and social impact analyses. Technical governance refers to the structures and procedures that make technology's activities responsible, transparent, and ethical by design, and then the technological audits come into play. Some of the technical aspects include ensuring robustness, bias, and explainability. Programs should be safe and protected, with no vulnerabilities to manipulation or compromise, including the training data. Systems should utilize training data and models that adjust for bias in data to prevent unfair treatment of particular groups. Tainted or skewed instances, restricted features, sample size difference, and proxies to protected characteristics are common causes of bias. Algorithms should make choices or provide recommendations that users and developers can understand. Individual and global explanations, model-agnostic, and model-specific interpretations are all crucial approaches in this area. Also, preventive steps and processes should be put in place to avoid potential damages. This preventive strategy necessitates anticipating hazards to reduce the likelihood of them happening and to minimize the effect if they do occur. Technical audits should be carried out throughout the development phase as well as during live monitoring. Impact evaluations are carried out before deployment and are used to develop mitigating measures. Although there will still be unknown risks, such activities can minimize the risk. Certification is a component of the risk management process that verifies that a system, method, or organization meets a set of requirements through initial or continuous audits. Certification is a final confirmation that may be obtained by presenting proof and demonstrating that a system, method, or organization has met the established criteria.


\section{Trustworthy AI}
Explainability, fairness, privacy, governance, and robustness are currently hot topics among researchers and adopters of AI, and they are grouped under the umbrella term "Trustworthy AI" \cite{brundage2020toward}. From an engineering standpoint, a real-time application of Trustworthy AI shall focus on four key elements: Accountability and privacy, discrimination and bias, explainability and interpretability, and robustness and performance.

\subsection{Accountability and Privacy}
Accountability and privacy connects to the principle of damage prevention. Customized data governance is required, which includes the quality and integrity of the data utilized, its relevance in the area where the algorithm will be employed, access procedures, and the capacity to handle data in a way that respects privacy \cite{de2020overview}. These problems may divide into two categories: (i) Privacy and data protection: Lifetime of a system must guarantee, privacy and data protection. It covers both the user information and the information produced about them through their contact with the system. Finally, data access procedures should be established, defining who has access to data and under what conditions. Data Protection Impact Assessment is the standard method for assessing risks \cite{lee2020innovating}. (ii) Model inferences: Any system's security assess in terms of the adversary objectives and capabilities it intend to counter. Inferring model parameters and creating "knock-off" versions of them is the primary attack vector in this component. To determine vulnerability, the auditor may use methods to extract a (near-)equivalent copy or steal part of an algorithm's functionality.

\subsection{Discrimination and Bias}
Multiple forms of bias exist in AI and ML, explaining how an automated decision-making process may become unjust. Due to human and social prejudices, every machine learning system retains the bias present in tainted data. Historical observations, such as police records, support previous predictions, resulting in a discriminative outcome. Under-sampled data from minority groups lead to unreliable consequences due to induced biases. To detect and reduce decision-making bias, we must first distinguish between personal and collective fairness. (i) Personal Fairness: tries to treat comparable people in the same way. (ii) Collective Fairness: divides the data into different groups based on protected characteristics and attempts to achieve equality to some degree across groups. It is also feasible to differentiate between equality of opportunity and outcome within the Collective Fairness. For instance, it is using the SAT score as a criterion for predicting college achievement. Note that fairness may be understood quite differently in various settings and nations; thus, a single implementation of a particular algorithm might run against many distinct fairness assessment obstacles. Finally, it is worth emphasizing that it is theoretically impossible to create an algorithm that meets all acceptable criteria of both a "fair" or "unbiased" engine at the same time.

\subsection{Explainability and Interpretability }
Explainability and interpretability are frequently used alternatively in the context of AI and ML. Interpretability is the degree of cause and effect of a system, and the extent to which an observer can anticipate what will happen for a particular set of input or algorithm parameters. Explainability refers to how easily an AI/ML system's explains the underlying mechanics. Interpretability is the ability to comprehend the mechanics of an algorithm, and explainability refers to the ability to describe what is going on in an algorithm. Building and sustaining users' confidence in automated decision-making systems requires giving clear and relevant explanations. Procedures must be transparent, system capabilities and purposes public disclosure, and choices must be explainable to people directly and indirectly impacted, to the degree feasible. A transparent system also helps the developer by allowing them to "debug" it, expose unjust choices, and gain information. Possible solutions to incorporate explainability and interpretability into AI/ML algorithms are classified as intrinsic and model agnostic approaches \cite{ehsan2021expanding}. In an intrinsic approach, a model is created and developed so that it is completely transparent and understandable by design with model-specific explainability. An extra explainability method does not need to be applied to the model to explain its functioning and outputs completely. In the model-agnostic approach, explainability is gained by applying mathematical methods to the findings of any algorithm, even extremely complicated and opaque models, to understand the decision factors for such models. It's essential to keep in mind that the explainability criteria for various locations and use cases may vary. A single method may not be appropriate in all situations when an algorithm is used.

\subsection{Robustness and Performance}
Algorithm Performance and Robustness refers to how well an algorithm can be considered safe and secure, not susceptible to tampering or compromising its trained data. Like other software systems, AI systems should be secured against vulnerabilities that may enable adversaries to exploit them, such as data poisoning, model leaking, or infrastructural facilities, both software and hardware.
This idea is connected to the mathematical notion of Adversarial Robustness \cite{carlini2019evaluating}, which asks how the algorithm would have fared in the worst-case situation. AI systems should include protections that allow for a backup strategy in the event of a malfunction.
In addition, the degree of safety precautions needed is determined by the size of the danger presented by an AI system.
This idea is closely linked to Verification, which implies, in general words, whether the method adheres to the issue requirements and restrictions. AI system's capacity to make accurate judgments, such as properly categorizing data into appropriate categories, or making correct forecasts, suggestions, or choices based on data or models, is referred to as Accuracy of a model. Accuracy as a broad notion may be measured by calculating Expected Generalization Performance, which implies that the issue of how well the algorithm works, in general, can be answered. A reliable AI system operates well with a variety of inputs and in various circumstances. At the same time, reproducibility refers to whether or not an AI experiment behaves the same when repeated under the same conditions.

\section{AI Algorithm - Quid Pro Quo}
In the nascent field of Trustworthy AI, no one size fits all solution, however just the trade-offs to be handled. Although the practicalities of trade-off analysis need context, broad investigations, road maps, and recommendations may still be given and implemented. Interpretability vs. Accuracy trade-off is often referred to as Explainability vs. Performance trade-off. It seems to be quite realistic at first glance; however, such portrayal is very controversial \cite{smietanka2021algorithms}. It is possible that a Linear model is the most accurate model but that the explainability of the model is significantly lowered owing to extensive pre-processing. Explainability vs. Fairness trade-off inclines towards improving a system's explainability to achieve more transparency in its usage and serves as a positive motivator for all of its users and designers to reveal underlying prejudice and discrimination. Fairness vs. Robustness is another well-studied trade-off for bias vs. performance. Fairness vs. Bias trade-off is another crucial factor to be considered while auditing an algorithm. In terms of privacy, the closer a system is to anonymity, especially in personal data, the more 'private' it is considered to be.
On the other hand, in the case of fairness, the issue is that systems function similarly for all protected characteristics. As a result, systems must be as accessible as possible to ensure fairness. The conflict between privacy and fairness emerges, with a higher privacy phase likely to come at the expense of concerns about justice.

Despite the importance of trade-off analysis, it should be emphasized that bringing all of these areas together is frequently difficult and not always desired. Trade-offs should be seen as a means of determining an operational profile that is compatible with the application's requirements, rather than an abstract objective that must be met to create a sense of completeness \cite{longo2020explainable}. One of the most difficult tasks is determining which risks should be prioritized and quantified. This is handled on a case-by-case basis, but a road map or toolkit may be created to assist business users and developers with the appropriate recommendations and areas to concentrate on, such as the following:
\begin{itemize}
    \item Performance and Robustness, such as when an algorithm's statistical accuracy or brittleness may result in financial and reputational harm.
    \item When there is a lack of comprehension of the choices being taken, recommendations being offered, or remedies being sought, interpretability and explainability are required.
    \item When the risk of intellectual property or private information being leaked is a real probability. Depending on the phase, the information given, and the kind of project involved, monitor metrics and suggest actions.
    \item Provide suggestions for helpful tools and methods to include in the development/procurement process so that risks may be minimized and avoided.
    \item Request information on performance, bias, and other metrics throughout the deployment phase to ensure that the risks are under control.
\end{itemize}

\section{Conclusion}
This report is an initial step toward understanding the main components that makeup AI Algorithm Auditing. We aim to initiate a discussion in this new field of study and practice and do so with a solid collection of topics, methods, and techniques. The effect of ideas like accountability, fairness, and transparency on design decisions, algorithms to be utilized, delivery methods, and physical infrastructure is not simple to translate into engineering practice. It necessitates a complete integration of governance structures as well as real-time algorithm audits. We anticipate that with the emergence of new sectors utilizing AI, auditing and assurance of data and algorithms will become crucial.

\section{Acknowledgments}
This research was funded by (a)University of Central Florida provost scholarship for joint research with National Academy members. (b)National Science Foundation on I-Corps program. Disrupting Legal Tech: Using Artificial Social Intelligence for Legal Team Assistants, Award Number:1933260. 

\bibliographystyle{splncs04}
\bibliography{bibliography}

\end{document}